# Measurement of Laterally Induced Optical Forces at the Nanoscale

*Fei Huang[1], Venkata Ananth Tamma[2] and H. Kumar Wickramasinghe [1*]*

[1] *Department of Electrical Engineering and Computer Science, 142 Engineering Tower, University of California, Irvine, USA*
[2]*CaSTL Center, Department of Chemistry, University of California, Irvine, USA*

*Corresponding author: hkwick@uci.edu

***Abstract***: We demonstrate the measurement of laterally induced optical forces using an Atomic Force Microscope (AFM). The lateral electric field distribution between a gold coated AFM probe and a nano-aperture in a gold film is mapped by measuring the lateral optical force between the apex of the AFM probe and the nano-aperture. Torsional eigen-modes of an AFM cantilever probe were used to detect the laterally induced optical forces. We engineered the cantilever shape using a focused ion beam to enhance the torsional eigen-mode resonance. The measured lateral optical force agrees well with simulations. This technique can be extended to simultaneously detect both lateral and longitudinal optical forces at the nanoscale by using an AFM cantilever as a multi-channel detector. This will enable simultaneous Photon Induced Force Microscopy (PIFM) detection of molecular responses with different incident field polarizations. The technique can be implemented on both cantilever and tuning fork based AFM's.

Lateral force AFM is a technique that is primarily used to differentiate nanoscale surface properties [1], [2]. In Lateral force AFM, the frictional forces between the tip and sample creates a torsion of the cantilever which in turn is a function of the surface properties, and leads to chemical force microscopy [3], [4]. In this letter we demonstrate the detection of lateral optical forces using the torsion mode of cantilever thereby enhancing the capability of lateral force AFM to detect optical forces at the nanoscale. Photon Induced Force Microscopy (PIFM) is a promising new technique to study linear and non-linear optical properties measured only using photon induced forces [5] - [9]. PIFM uses an Atomic Force Microscope (AFM) to measure the optical forces between an optically induced dipole in the sample under measurement and another optically induced dipole formed at the tip of the gold coated AFM probe. Previously, PIFM was introduced to detect and image linear molecular resonances at nanometer level [5], [7], [8] and perform non-linear imaging and spectroscopy at the nanoscale [6], [9] as well as time-resolved imaging of non-linear optical properties [9] of molecules. Indeed, PIFM has been used to image molecular resonances over a wide range of wavelengths from the visible to mid-IR wavelength regimes [10]. In addition, optical forces between a gold coated AFM tip and its image on a glass substrate was used to image the focal field distributions of tightly focused laser beams with different polarizations [7]. The response of the tip to different polarizations was used to estimate the aspect ratio of the AFM probe tip making it a useful technique to estimate the quality of probes for sensitive experiments such as Tip Enhanced Raman Spectroscopy (TERS) [7]. The previous works used the AFM in the attractive mode to measure the component of optical force, along the tip axis.

In this paper, we demonstrate the use of PIFM to measure the lateral optical force (perpendicular to the tip axis) between a gold coated AFM probe and a nano-aperture in a gold film both illuminated by a focused optical field. To measure the lateral component (along *x*-axis, refer to Fig. 1 for definition of directions) of the total optical force, we used the first torsional resonance of the AFM cantilever. Topography was obtained using the first flexural resonance of the AFM tip. The second flexural resonance was used to detect the longitudinal (*z*-direction) component of the optical force. The total force experienced by the AFM probe tip is $\vec{F}_{tot} = \langle \vec{F}_{opt} \rangle + \vec{F}_{int}$, where, $\vec{F}_{int}$ is the total tip-sample interaction force consisting of all chemical, Casimir, meniscus and van der Waals forces. Here, $\langle \vec{F}_{opt} \rangle$ is the total time-averaged optical force on the AFM probe tip due to its interaction with the incident field and induced dipole on the molecule [7]. It can be expressed along different directions as

$\langle \vec{F}_{opt} \rangle = \langle F_{opt,x} \rangle \hat{x} + \langle F_{opt,y} \rangle \hat{y} + \langle F_{opt,z} \rangle \hat{z}$. Previously, results [5]-[9] were obtained by using only the flexural modes of the AFM cantilever and therefore $\langle F_{opt,z} \rangle$ is the only component of the total force that was detected by the AFM cantilever. However, as shown schematically in Fig. 1 (a), both the AFM cantilever torsional resonance and tuning fork AFM in shear force mode could be used to detect the *x*-component $\langle F_{opt,x} \rangle$ of the total time averaged optical force. Also, by simultaneously recording data from both the flexural and torsional modes, the different optical force components can be measured.

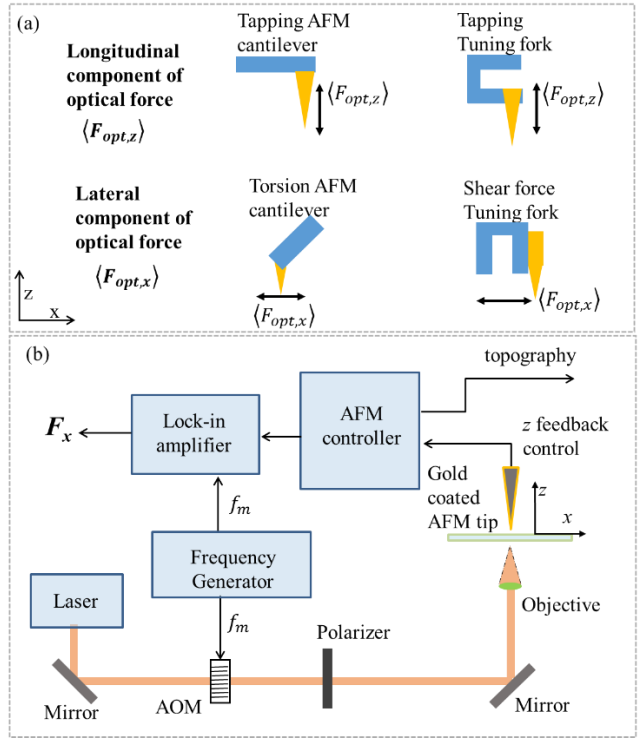

Fig. 1 (a) Schematic of different AFM modes include the tapping mode AFM cantilever which detects optical force in *z* direction, AFM cantilever in torsional mode detects optical force in *x* direction, tuning fork AFM in shear force mode detects optical force in *x* direction while tuning fork AFM in tapping mode detects optical force in *z* direction and (b) Schematic of experiment setup used to measure the lateral optical force $F_x$

The experiment is built around a commercial AFM (Veeco Instruments Caliber) operating in the attractive mode and is detailed in Fig. 1 (b). The optical beam under measurement is modulated by an Acousto-Optic Modulators (AOM). The modulation frequency $f_m$ was chosen depending on whether the lateral or longitudinal component of the total optical force was under measurement. To measure the lateral component (along *x*-axis) of the total optical force, the optical beam was modulated at the frequency of the first torsional resonance ($f_{0,Tor}$) of the AFM cantilever. To measure the longitudinal (along *z*-axis) component of the total optical force, the optical beam was modulated at the frequency of the second flexure mode ($f_{1,Flex}$) of the AFM cantilever. The modulated light was then focused on the sample fabricated on a 0.16 mm thick cleaned glass cover slip using an oil immersion objective (Olympus PlanApo 100x) with *NA* = 1.45. The modulated optical focal field results in a corresponding modulation of the optical force induced between the AFM probe tip and the sample. The detected optical force is recovered by using a lock-in amplifier with a reference at the modulation frequency. The gold coated AFM cantilever probes were prepared by sputter coating (South Bay technology) commercial bare Silicon AFM cantilever probes (AppNano Forta) with 25 nm gold on a 2 nm chromium adhesion layer.

Since the measured optical force signal $F_x$ is proportional to the AFM cantilever deflection at a given frequency, the measured optical force is expressed in terms of the cantilever spring constant and quality factor $Q$ for a particular mode as $F_i = \langle F_{opt,i} \rangle Q_m / k_m$, where, $i = x, z$ and $m$ corresponds to the second flexural mode for $i = z$ and to the first torsional mode for $i = x$. The measured optical force signal is amplified by the high quality factor of the particular mode being detected. In order to increase the detected signal, we need to reduce the stiffness of the cantilever and increase its Q. The spring constant of the torsional mode is typically very large [11], [12]. To reduce the spring constant of AFM cantilevers used in this work were modified them following the technique described in [11],[12]. We first measured the resonance frequencies and quality factors of the AFM cantilevers in free space. . Resonance frequencies of the first flexural, second flexural and first torsional eigen-modes were verified using the structural mechanics module of the commercial finite-element code COMSOL Multiphysics. Hinges were cut into the cantilever using Focused Ion Beam (FIB) milling as shown in the scanning electron micrographs (SEM) shown in Fig. 2 (b). The inset in Fig. 2 (b) shows a zoomed-in SEM of the hinge. For reference, a SEM of the original cantilever without hinges is shown in Fig. 2 (a). The resonance frequency values of the cantilever eigen-modes after FIB milling were re-measured and were computationally verified using COMSOL Multiphysics.

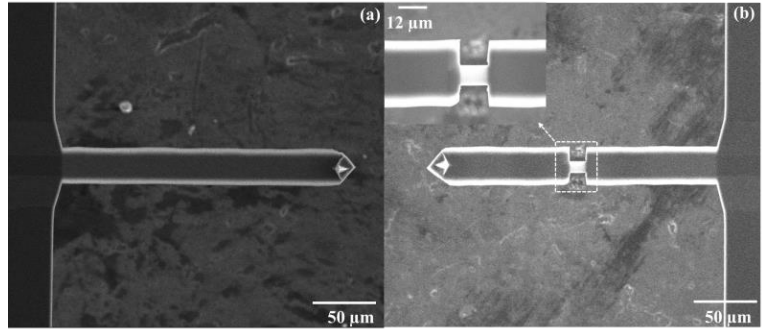

Fig. 2 (a) Scanning electron micrograph of AFM cantilever without hinge and (b) scanning electron micrograph of cantilever with hinge with inset showing zoomed-in scanning electron micrograph of the hinge.

Two different AFM cantilevers from the same wafer were used to obtain the results presented here. Tip. 1 was used to produce the results in Fig. 3 (a) and (b) while Tip 2 was used to obtain results in Fig. 3 (c) and (d). Both the tips had similar values for the resonance frequencies of the first flexural, second flexural and first torsional eigen-modes. Before FIB milling, the measured resonance frequencies of the first flexural, second flexural and first torsional eigen-modes for both tips were =around 58 kHz, 380 kHz and 850 kHz, respectively. After FIB milling, the measured resonance frequencies of the first flexural, second flexural and first torsional eigen-modes for both tips were around 51 kHz, 235 kHz and 735 kHz, respectively.

We mapped the optical force distribution between a gold coated AFM probe and single nano-aperture diameter 300 nm, in a 50 nm thick gold film deposited on a 0.16 mm thick clean glass coverslip. The gold film was sputter coated (South Bay technology) on the coverslip with a 2 nm chromium adhesion layer. Nano-apertures of diameter 300 nm were then prepared by FIB milling the gold film down to the glass ensuring no residual gold was left within the the nano-aperture. The nano-apertures were placed in the focal spot of a focused laser beam with a wavelength of 680 nm. The average power of the laser beam was 100 $\mu$W. We first mapped the optical force between a gold coated AFM probe and a single nano-aperture of diameter 300 nm for a linearly polarized (along x-direction) laser beam focused on the nano-aperture. Since tightly focused laser beams with linear polarization have both longitudinal (z-directed) and lateral (x-directed) electric fields in the focal plane, we reduced the diameter of the incident laser beam to 33 % of the diameter of the focusing objective back aperture. This reduced the effective *NA* of the focusing objective ensuring that only lateral (*x*-direction) electric field component in the focused laser spot was dominant. The simultaneously recorded topography and normalized lateral optical force are plotted in Fig.3 (a) and (b), respectively. In Fig. 3 (c) and (d), we present the simultaneously recorded topography and normalized longitudinal (*z*-direction) optical force distributions, respectively, of a gold coated AFM probe interacting with a single nano-aperture of diameter 300 nm but illuminated by incident tightly focused light with radial polarization. The nano-aperture was positioned to be within the central spot of the tightly focused radially polarized beam having purely longitudinal electric fields ($E_z$).

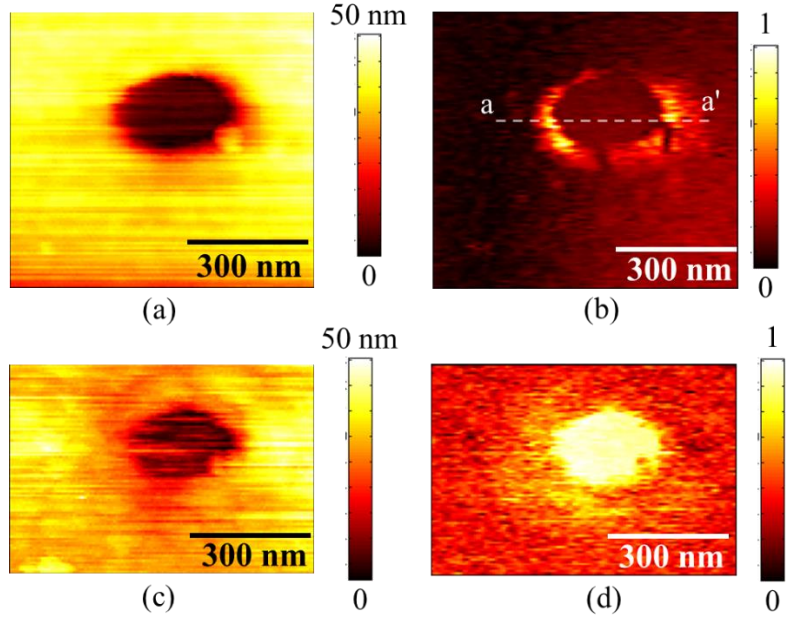

Fig. 3. Spatial distributions of (a) topography and (b) normalized lateral optical image force measured experimentally at 680 nm with linearly polarized input light (c) topography and (d) normalized longitudinal optical image force measured experimentally at 680 nm with radially polarized input light

To understand the spatial distribution of normalized optical force plotted in Fig. 3 (b) we computed the electric field distribution around a single nano-aperture of diameter 300 nm in a 50 nm thick gold film with linearly polarized plane wave input using the RF module of COMSOL Multiphysics. A cross-section plot of the spatial distribution of the normalized lateral (*x*-directed) electric field component around a single nano-aperture is shown in Fig. 4 (a) (the white colored arrows indicate the direction of electric fields). In the COMSOL simulations, the edges of the nano-apertures were filleted with a radius of curvature 10 nm. We note the thickness of gold film is greater than the skin depth of gold at 680 nm. The optical constants of gold and silicon used in all calculations in the work were obtained from [14], [15]. In Fig. 4 (a), we observe a strong field enhancement near the edges of the nano-aperture indicating that the edges will have the strongest interaction with the apex of the gold coated AFM tip due to lateral optical force. We performed further COMSOL simulations of a gold coated AFM probe interacting with the nano-aperture and compared the line trace of lateral optical force obtained experimentally with the simulations in Fig. 4 (b). The results of the simulations are in general agreement with experiment. The slightly broadened experimental data can be explained by considering the shape of the AFM probe. Further optimization of the AFM cantilever to enhance the sensitivity of the torsional mode will help improve the signal to noise ratio. The results in Fig. 4 (d) can be understood by noting that the full Width Half Maximum of the central spot of a tightly focused radial beam, with purely longitudinal (*z*-direction) electric fields, at wavelength 680 nm is roughly 360 nm [7], [13] and completely illuminates the nano-aperture of diameter 300 nm precisely aligned to the central spot. It leads to the longitudinal force map of the nano-aperture as observed in Fig. 4 (d).

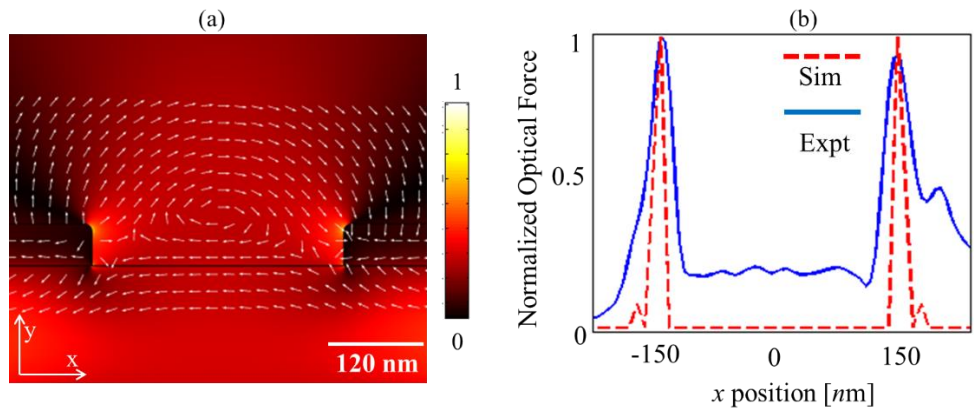

Fig. 4. Numerical calculations of normalized lateral (*x*-direction) electric fields around a single nano-aperture of diameter 300 nm in 50 nm thick gold film. The white colored arrows indicate the direction of electric fields. (b) Comparison of line traces of normalized optical force obtained experimentally (line a-a' in Fig. 3 (b)) and COMSOL simulations showing good agreement

This technique can be extended for simultaneous detection of lateral and longitudinal optical forces. To achieve this, the AFM cantilever will be used as a multi-channel nanoscale force detector with each detection channel corresponding to a different eigen-mode of the cantilever [16]. For example, by modulating the incident laser beam at two frequencies $f_1$ (second flexure resonance) and $f_2$ (first torsional resonance) using two acousto-optic modulators, the cantilever will respond to these modulated fields at its respective resonant frequencies (second flexure and first torsional) leading to simultaneous detection of both longitudinal and lateral optical forces. By using the AFM cantilever as a multi-channel detector to measure optical forces along different directions simultaneously, this technique can be generalized to measure the tensorial components of molecular resonances. We note that the technique is completely general and can be implemented using both cantilever and tuning fork based AFM's. In conclusion, we have demonstrated the detection of lateral optical forces using the PIFM technique. We used the torsional eigen modes of a cantilever to measure the lateral optical force distribution between a gold coated AFM probe and a nano-aperture of 300 nm diameter in a gold film of thickness 50 nm. The resonance frequency of the torsional eigen-mode of the AFM cantilever was optimized by engineering the cantilever shape using a focused ion beam. The measured lateral optical force distribution agreed well with our simulations.

This work was supported by the NSF Center for Chemical Innovation, Chemistry at the Space-Time Limit (CaSTL) under Grant No. CHE-1414466. The authors thank Prof. Ara Apkarian, Prof. Eric Potma, Dr. Junghoon Jahng from the Department of Chemistry, University of California, Irvine, Dr.Jinwei Zeng, Yinglei Tao from the Department of Electrical Engineering, University of California, Irvine and Dr. Jian-Guo Zheng in in the UC Irvine Materials Research Institute (IMRI) for helpful discussions. We acknowledge South Bay Technology for allowing us to use its Ion Beam Sputtering/ Etching (IBS/e) System) in the preparation of AFM cantilever probes in the UC Irvine Materials Research Institute (IMRI)